\documentclass[epj]{svjour}
\usepackage{graphics}

\newcommand{\SKIP}[1]{}
\newcommand{\eps}{\varepsilon}

\begin{document}       %=================================================
\title{Exploring high-density baryonic matter:\\ Maximum freeze-out density}

\titlerunning{Freezeout density}

%\subtitle{Do you have a subtitle?\\ If so, write it here}
\author{J{\o}rgen Randrup\inst{1} \and Jean Cleymans \inst{2}
}                     % Do not remove

\institute{Nuclear Science Division, %LBNL, 
Lawrence Berkeley National Laboratory,
Berkeley, California 94720, USA
\and
%%%%%%%%%%%%%%%%%changed affiliation%%%%%%%%%%%%%%%%%%%%%%%%%
UCT-CERN Research Centre and Department of Physics, University of Cape Town, South Africa}
%%%%%%%%%%%%%%%%%end changed affiliation%%%%%%%%%%%%%%%%%%%%%%%%%
%
\date{Received: date / Revised version: date}
% The correct dates will be entered by Springer
%
\abstract{
The hadronic freeze-out line is calculated in terms of 
the net baryon density and the energy density
instead  of the usual $T$ and $\mu_B$.
This analysis makes it apparent that the freeze-out density exhibits
a maximum as the collision energy is varied.
This maximum freeze-out density has $\mu_B=400-500\,{\rm MeV}$,
which is above the critical value, and it is reached
for a fixed-target bombarding energy of $20-30\,{\rm GeV/N}$ 
well within the parameters of the proposed NICA collider facility.
 % end of PACS codes
} %end of abstract

\maketitle
%
%------------------------------------------------------------------------
%\section{Introduction}
%\label{intro}

Over the past decade a striking regularity has been established
in relativistic nuclear collisions: From the lowest beam energies to the highest,
the yields of various hadronic species are consistent with the assumption of
chemical equilibrium \cite{Becattini,wheatonphd,marek,jaakko,andronic}. 
Analyses of the experimentally obtained hadronic yield ratios
at a variety of collision energies have shown that the data
can be well reproduced within the conceptually simple statistical model
that describes an ideal hadron resonance gas in statistical equilibrium.
Furthermore, the extracted freeze-out values of the temperature $T$ and
the baryon chemical potential $\mu_B$ exhibit a smooth and monotonic
dependence on the collision energy and can be simply parametrized.

The collision energy thus plays a determining role for the thermodynamic
 properties of the freeze-out state in relativistic nuclear collisions.
However, there is no simple relationship between the collision energy 
and the freeze-out value of the (net) baryon {\em density}:
At low energies the freeze-out density increases with the energy,
whereas it decreases when the energy is high
due to the onset of nuclear transparency.
Thus there must exist a certain range of collision energies 
within which the freeze-out values of the net baryon density
displays a maximum.

The optimal collision energy leading to this highest freeze-out density 
was discussed in \cite{RandrupPRC74} 
on the basis of the up-to-date results on the properties of the final state.
It was pointed out there that 
since neither $\mu_B$ nor $T$ is subject to a conservation law
they may be less suitable in a dynamical context.
Furthermore, when a first-order phase transition is present,
they become multivalued functions of the basic mechanical variables 
$\rho_B$ (net baryon density) and $\eps$ (energy density)
inside the mixed-phase region.
It is therefore of interest to reexpress the thermodynamic variables
in terms of those mechanical densities.
Accordingly,
we considered in \cite{RandrupPRC74} how the freeze-out line appears
when represented in terms of the basic baryon and energy densities,
rather than chemical potential and temperature.

We present here updated and refined results
using the latest version of THERMUS~\cite{THERMUS} 
together with an updated input of hadronic resonances~\cite{PDG}.
The calculations in \cite{RandrupPRC74} were made without including
any effects of the hadronic interactions at freeze-out.
Because those may significantly affect the results
(but not the qualitative features),
we also present results that take approximate account of the interactions
by means of an excluded volume.
The hard core radius for the excluded volume was chosen to be 
$c=0.3$~fm for all hadrons. The results for $c=0.6$~fm are also shown. 
A comprehensive analysis of the effects of a hard-core radius 
was first presented in~\cite{gorenstein}. 

In \cite{RandrupPRC74}, we presented the freeze-out line
in terms of the net baryon density $\rho_B$ and the total energy density $\eps$.
We employ below, in Fig.\ 1, the corresponding $(\rho_B,\eps^*)$ representation,
where the ``excitation energy density'' $\eps^*\equiv\eps-m_N\rho_B$ 
is the energy density above the minimum value $m_N\rho_B$ 
dictated by the specified net baryon density.
Thus $\eps^*$ has both compressional and thermal contributions.

\begin{figure}[t]      %---------------------------------------------------
\resizebox{0.5\textwidth}{!}{%
  \includegraphics{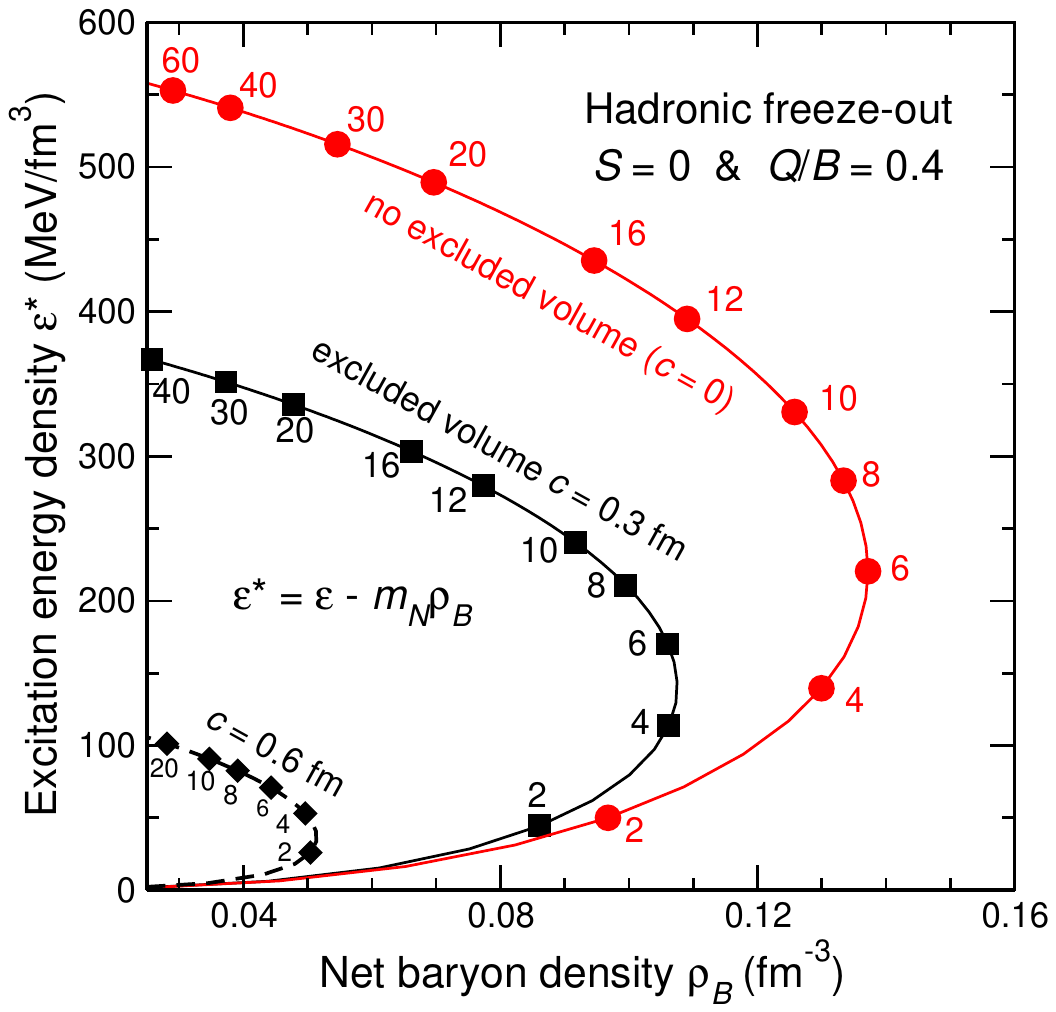}
}
\caption{The hadronic freeze-out line in the $\rho_B-\eps^*$ phase plane 
as obtained from the values of $\mu_B$ and $T$
 that have been extracted from the experimental data~\cite{wheatonphd}.
The calculation employs values of $\mu_Q$ and $\mu_S$ that ensure 
$\langle S\rangle=0$ and $\langle Q\rangle=0.4\langle B\rangle$
for each value of $\mu_B$.  
The solid circles correspond to the results of \cite{RandrupPRC74}
that were obtained without using any excluded volume,
while the solid squares show the corresponding results
calculated with an excluded volume having a radius of $c=0.3$~fm. 
The diamonds were obtained with $c=0.6$~fm.
Each point is labeled by the collider beam energy (in GeV/$N$)
for which the particular freeze-out conditions are expected.}
\label{f:rho-eps}     
\end{figure}	       %---------------------------------------------------

We also show, in Fig.\ 2, the corresponding $(\rho_B,T)$ diagram,
because the freeze-out temperature $T$ is perhaps more easily grasped
than the excitation energy density $\eps^*$.

\begin{figure}[t]      %---------------------------------------------------
\resizebox{0.5\textwidth}{!}{%
  \includegraphics{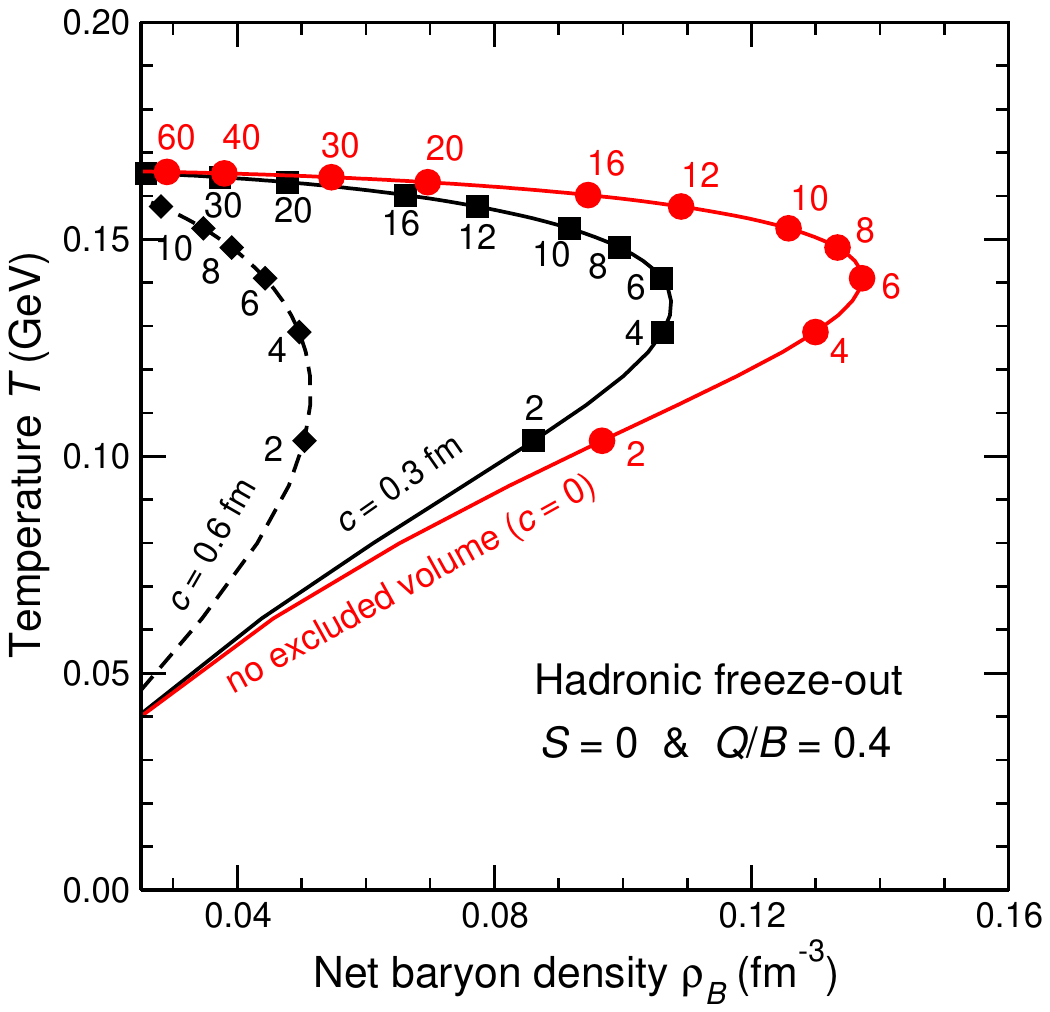}
}
\caption{The hadronic freeze-out line in the $\rho_B-T$ phase plane 
as obtained from the values of $\mu_B$ and $T$
 that have been extracted from the experimental data~\cite{wheatonphd}.
The calculation employs values of $\mu_Q$ and $\mu_S$ that ensure 
$\langle S\rangle=0$ and $\langle Q\rangle=0.4\langle B\rangle$
for each value of $\mu_B$.  
The solid circles correspond to the results of \cite{RandrupPRC74}
that were obtained without using any excluded volume,
while the solid squares show the corresponding results
calculated with an excluded volume having a radius of $c=0.3$~fm. 
The diamonds were obtained with $c=0.6$~fm.
Each point is labeled by the collider beam energy (in GeV/$N$)
for which the particular freeze-out conditions are expected.}
\label{f:rho-T}     
\end{figure}	       %---------------------------------------------------

These novel representations of the freeze-out line bring out very clearly 
that there is a maximum value of the net baryon density:
At the highest collision energies, 
freeze-out occurs for a negligible value of $\rho_B$
and at an energy density of nearly one half $\rm GeV/fm^3$;
then, in the range of $\mu_B=400-500\,{\rm MeV}$
(and a temperature of $T=140-130\,{\rm MeV}$),
the freeze-out line exhibits a backbend and approaches the origin.
Thus, the net baryon density at freeze-out has a maximum value
which amounts to about three quarters of the familiar 
nuclear saturation density of $\rho_0\approx0.16\,{\rm fm}^{-3}$.  

The fact that the freeze-out value of the net baryon density
exhibits a maximum as the collision energy is being scanned
suggests that the corresponding collision energy (range) is optimal
for the exploration of compressed baryonic matter.
Based on the present calculations,
this infered optimal collision energy is 
$\sqrt{s_{NN}}=5.6-7.8\,{\rm GeV}$ for a collider 
(such as RHIC at BNL or NICA at JINR).
The corresponding optimal beam kinetic energy is
$15-30\,{\rm GeV}$ per nucleon for a fixed-target configuration 
(such as FAIR at GSI).

The results presented here should provide valuable guidance
for establishing the desired capabilities of the planned NICA at JINR.
In particular, our results suggest that freeze-out densities
all the way up to the maximum value could be explored
 at a collider facility delivering beam kinetic energies of 
up to $\approx$2.4~GeV per nucleon.

%------------------------------------------------------------------------
\begin{acknowledgement}
This work was supported the Office of Nuclear Physics 
in the U.S.\ Department of Energy's Office of Science 
under Contract No.\ DE-AC02-05CH11231 (JR).
\end{acknowledgement}
~\\[8ex]

		  %--------------------------------------

\end{document}